\documentclass[useAMS,usenatbib]{mn2e}
\usepackage{graphicx}
\usepackage{amssymb}
\usepackage{subfigure}
\usepackage[utf8]{inputenc}
\usepackage[english]{babel}
\usepackage{amsfonts}
\usepackage{amsbsy}
\usepackage[usenames]{color}
\usepackage{multirow}
\usepackage{rotating}
\usepackage{slashbox}
\usepackage{ctable}
\usepackage{ulem}
\usepackage{subfigure}
\usepackage{amsmath}

\newcommand{\be}{\begin{equation}}
\newcommand{\ee}{\end{equation}}
\newcommand{\bea}{\begin{eqnarray}}
\newcommand{\eea}{\end{eqnarray}}

\newcommand{\ben}{\begin{eqnarray}}
\newcommand{\een}{\end{eqnarray}}

\begin{document}
\title[WSLAP+]{Enabling Non-Parametric Strong Lensing Models to Derive Reliable 
Cluster Mass Distributions. WSLAP+} 

%Improved lensing reconstruction with WSLAP+]{WSLAP+: Improved lensing reconstruction with a non-parametric code}

\author[]{Irene Sendra$^{1}$, Jose M. Diego$^{2}$, Tom Broadhurst$^{1,3}$
Ruth Lazkoz$^{1}$\\
$^{1}$Department of Theoretical Physics, University of the Basque Country UPV/EHU, 48080 Bilbao, Spain \\
$^{2}$IFCA, Instituto de F\'isica de Cantabria (UC-CSIC), Av. de Los Castros s/n, 39005 Santander, Spain\\
$^{3}$IKERBASQUE, Basque Foundation for Science, Alameda Urquijo, 36-5 48008 Bilbao Spain}
\date{\today}
%\pacs{}
\pagerange{\pageref{firstpage}--\pageref{lastpage}}
\maketitle
%%%%%%%%%%%%%%%%%%%%%%%%%%%%%%%%%%%%%%%%%%%%%%%%%%%%%%%%%%%%%%%%%%%%%%%%%%%%%%%  
\label{firstpage}
\begin{abstract}

%We show that non-parametric strong-lensing
%methods can reliably determine the general distribution of cluster
%dark matter by incorporating a simple prior for the member
%galaxies

 In the strong lensing regime non-parametric lens models struggle to
 achieve sufficient angular resolution for a meaningful derivation of
 the central cluster mass distribution. The problem lies mainly with
 cluster members which perturb lensed images and generate additional
 images, requiring high resolution modelling, even though we mainly
 wish to understand the relatively smooth cluster component. In
 practice the required resolution for a fully non-parametric mass map
 is not achievable because the separation between lensed images is
 several times larger than the deflection angles by member galaxies,
 even for the most impressive data. Here we bypass this limitation by
 incorporating a simple physical prior for member galaxies, using
 their observed positions and their luminosity scaled masses. This
 high frequency contribution is added to a relatively coarse Gaussian
 pixel grid used to model the more smoothly varying cluster mass
 distribution, extending our established WSLAP code \citep{Diego2007}. We test 
this new code (WSLAP+) with an empirical
 simulation based on A1689, using all the pixels belonging to
 multiply-lensed images and the observed member galaxies. Dealing with
 the cluster members this way leads to stable convergent solutions,
 without resorting to regularization, reproducing well smooth input
 cluster distributions and substructures. We highlight the ability of
 this method to recover ``dark'' sub-components and other differences
 between the distributions of cluster mass and member galaxies. Such
 anomalies can provide clues to the nature of invisible dark matter,
 but are hard to discover using parametrized models where
 substructures are  modelled on the basis of the visible data. With our
 increased resolution and stability we show, for the first time, that
 non-parametric models can be made sufficiently precise to locate
 multiply-lensed systems, thereby achieving fully self-consistent
 solutions without reliance on input systems from less objective
 means.

\end{abstract}

%%%%%%%%%%%%%%%%%%%%%%%%%%%%%%%%%%%%%%%%%%%%%%%%%%%%%%%%%%%%%%%%%%%%%%%%%%%%%%%  

%%%%%%%%%%%%%%%%%%%%%%%%%%%%%%%%%%%%%%%%%%%%%%%%%%%%%%%%%%%%%%%%%%%%%%%%%%%%%%%    
%%%%%%%%%%%%%%%%%%%%%%%%%%%%%%%%%%%%%%%%%%%%%%%%%%%  
\section{Introduction}\label{section_introduction}  

  The distribution of mass within clusters is sensitive to the nature
  of dark matter and to the evolution of structure in general. In
  successful hierarchical models based on nonrelativistic and
  collisionless Cold Dark Matter (CDM) \citep{Peebles1984}, clusters
  accumulate from material gravitating towards the intersections of a
  filamentary network of structure, continuously merging with each
  other and increasing in mass. In this context simulations have shown
  that individual cluster mass profiles are well characterised in CDM
  dominated N-body simulations by the logarithmically steepening
  Navarro-Frenk-White (NFW) profile \citep{Navarro1997}, and also show
  a tendency towards a lower level of concentration with increasing
  cluster mass, the c-m relation, \citep{Bullock2001, Eke2001,
  Dolag2004, Maccio2008, Neto2007, Duffy2009, Zhao2009,
  Bhattacharya2011} reflecting the general later assembly of more
  massive structures, when the cosmic mean density is lower. Both of
  these predicted trends are now very well established by independent
  simulations, but with some interesting variations, mainly in the
  amplitude of the c-m relation \citep{Bhattacharya2011} that may require
  further clarification.

  Accurate and reliably constrained cluster mass profiles can now be
  measured by combining Strong (SL) and Weak Lensing (WL) information, providing
  full logarithmic radial coverage \citep{Broadhurst2005, Umetsu2008,
  Zitrin2009, Zitrin2010, Coe2011}. Rigorous comparisons with standard
  particle-CDM reveals that the shape of the profile follows closely
  the standard NFW profile for particle-CDM mass advocated to describe
  all halos formed in simulations of standard particle-CDM
  \citep{Broadhurst2005, Umetsu2010}. Curiously, however, the mass
  concentrations seem to be systematically larger than expected for
  the most massive clusters formed in the standard $\Lambda$CDM
  cosmological model, with approximately twice as much matter
  concentrated within the characteristic radius of the NFW profile
  \citep{Broadhurst2005, Umetsu2010}. This anomaly will be explored 
  in a more statistical sample from the CLASH survey (combining strong
  and weak lensing for full logarithmic coverage of the mass profile
  with first for an x-ray selected sample of 20 relaxed clusters
  \citep{Postman2011} for which the first results are quite intriguing
  \citep{Zitrin2010, Coe2011}. 

  Inherent triaxiality of dark matter halos can boost lensing based
  concentrations for clusters selected in the first place by their
  lensing strength \citep{Oguri2005}. By selecting according to other
  unrelated criteria this lensing bias may largely be avoided. The
  Hubble Treasury data for the CLASH program \citep{Postman2011} aims
  to establish representative equilibrium mass profiles for clusters
  selected by their X-ray properties, to be relaxed in appearance. The
%   initial results from this program combining weak and strong lensing 
  measurements are also in very good agreement with the NFW dominated
  CDM prediction \citep{Zitrin2010, Umetsu2011, Coe2011}
  but continue to lie tantalisingly above the concentration-mass
  relation predicted for halos formed late in the concordance $\Lambda$CDM
  cosmology. This surprising tendency for higher concentrations is
  controversial, although indications from other unbiased cluster
  samples with WL measurements also indicate higher cluster
  concentrations on average \citep{Oguri2009, Okabe2008},
  although these results are lacking the SL information for deriving
  the inner profile.

   Other dark matter related anomalies may have been found
  during cluster collisions, including complex merging cluster A2744
  \citep{Merten2011}, with evidence of anomalous density peaks
  of dark matter separated from galaxies and gas. Other such tentative
  claims have also been made in the case of merging cluster A520,
  where a central peak of dark matter is claimed without a
  corresponding enhancement in the number density of galaxies. In the
  case of the iconic Bullet-cluster, the large relative velocity
  inferred from the Mach cone of the bullet component, $4800$ $km/s$,
  \citep{Markevitch2004} is claimed to be very unlikely in the
  context of $\Lambda$CDM for which the maximum expected inferred initial
  impact velocity found in large simulations is around $\sim 1600$ $km/s$
  for pairs of halos exceeding $2\times10^{14}M_{\odot}$ 
  \citep{Mastropietro2008, Thompson2012}, Iliev et al. 2013 in preparation 
  but see also \citep{springel2007} for a possible explanation).

%   and with rapidly varying density fluctuations on scales of many Kpc
%   particularly when clusters are colliding, potentially observable as
%   deviations in the lensing deflection fields derived for clusters.

 This investigation is motivated empirically in view
 of the above imperfect agreement on cluster scales claimed between
 $\Lambda$CDM and the radial cluster mass profiles. We may seek variations on
 the standard CDM based cosmology. It may be argued that currently
 undetectable light axions or other low mass scalar field particles,
 are perhaps now better motivated as CDM candidates rather than the
 traditional super-symmetric CDM particle candidates that remain
 undiscovered to the highest energies reached to date. A
 characteristic feature of scalar fields is that the associated bosons
 can form a coherent BEC under suitable
 conditions of temperature and density, which can be initially met and
 maintained in the cosmological context \citep{Boehmer2007,Sikivie2009}. 
 Due to the low velocity dispersion, the growth of
 condensed structure should be very similar to standard particle-CDM
 on large scales \citep{Boehmer2007,Sikivie2009, Velten2012} as
 desired, but the macroscopic quantum wavelike behaviour may appear on
 smaller scales \citep{Choi2002, Woo2009, Gonzalez2011}, particularly
 when dark matter collides, with potentially interesting consequences
 for cluster lensing. In this context, the troublesome cores of dark
 matter dominated dwarf galaxies can be generated by setting the
 de Broglie wavelength to a scale of several $Kpc$, corresponding to a
 mass of $\sim 10^{-22}$ $eV$, so the uncertainty principle means matter
 cannot be confined within this radius \citep{Hu2000}.

   For this BEC form of CDM, full 3D simulations of the development of
   structure are computationally much more intensive than standard
   N-body simulations, but have recently begun, governed by a
   Schr\"odinger-Poisson equation \citep{Woo2009} to describe the
   balance between quantum pressure arising from the uncertainty
   principle offsetting the gravitational potential of the dark matter
   and with the corresponding particle mass set to the preferred value
   from dwarf galaxies cores. This simulation shows that the
   filamentary pattern and distribution of clusters is
   indistinguishable from regular simulations of particle-CDM, as
   expected \citep{Widrow1993}, but low mass galaxy
   halos are suppressed and large scale macroscopic quantum
   interference patterns are visible \citep{Woo2009} in the density
   distribution.  Comparison of our detailed cluster lensing mass
   distribution with predictions from this wave-like form of CDM \citep{Woo2009} 
   will be exciting to pursue and with our new lensing
   technique and can then be sensitively contrasted with the standard
   N-body representation of more massive particle-dark matter for which
   coherent wavelike behaviour is absent.

   We may now use gravitational lensing to search for dark matter
   anomalies with much increased precision as the data required to
   measure accurate mass distributions has leapt in quality over the
   past few years in both the strong and weak lensing regime. Many
   sets of multiple images are now very typically identified in deep
   multicolour Hubble data, where distinctive internal features can be
   recognised in the larger well resolved background galaxies
   \citep{Broadhurst2005, Umetsu2012, Zitrin2013}. To identify more typical, smaller and fainter multiply
   lensed sources it is necessary in practice to be guided by a lens
   model, as even for the best behaved clusters large perturbations
   from galaxy members locally distort one or more members of each set
   of multiple images so that the location of counter images cannot be
   guessed with any confidence and model inversion will fail. Without
   many complete sets of multiple-images spread over a range of
   redshift it is not possible to accurately constrain the inner mass
   profile of a cluster, sufficiently well to examine theoretical
   predictions.

  To take full advantage of this increased quality of data, many new
  approaches have been suggested to recover the surface mass
  distribution in both the weak and strong lensing regime, (see for
  instance \cite{Kaiser1993, Broadhurst1995, Kaiser1995,
  Schneider1994, Schneider1995, Seitz1995, Bartelmann1996, Taylor1998,
  Tyson1998, Bridle1998, Marshall2002}. In the best known case of
  A1689, over 100 multiply lensed images are reliably identified
  \citep{Broadhurst2005, Coe2010}, and over 50 are known in
  similar quality data for Cl0024+1654 \citep{Zitrin2009}, A1703 \citep{LimousinM.2008}, MACS0416-2403 \citep{Zitrin2013}, helped by the development of detailed
  parametric models, and in particular the simple method of \citep{Broadhurst2005}  where the cluster mass distribution is assumed to approximately
  trace the light, by first starting from the observed galaxy
  distribution and varying the coefficients of a low order 2D
  polynomial fit to the galaxy distribution to describe the general
  distribution of galaxy cluster mass, and in addition to this the member
  galaxy perturbations scaled by their luminosity, so that very few
  parameters are required to provide a fairly flexible model of the
  mass distribution, which can be used and refined in locating
  multiply lensed images.

   This relatively flexible method, although capable of locating many
  reliable multiple-images, is not precise enough to provide an
  exhaustive identification of all counter-images, particularly the
  numerous blue galaxies which are too ambiguous both morphologically
  and in terms of their estimated redshifts, and fundamentally this
  method is limited to self-consistency checks of models
  where mass traces light, as with standard CDM. To examine the data
  in detail for anomalous density fluctuations such as those that may
  be generated by wave-like CDM, we need the full
  model-independence that non-parametric strong lensing methods may
  provide. The increased number of strong lensing constraints
  available in deep space images encourages the use of non-parametric
  methods that make no assumption about the matter distribution. In
  previous work, we developed a non-parametric code (WSLAP) and
  demonstrated its performance first with simulated data, and later
  with the real data of A1689  \citep{Diego2005a, Diego2005b, Diego2007}. Our results were
  compared with those obtained using parametric methods
  \citep{Broadhurst2005} and found good agreement within the noise, in
  terms of the azimuthally averaged radial profile. However, the solution 
  obtained from WSLAP lacked the resolution of parametric methods limiting its 
  ability to predict new images that could be later confirmed with the data. 
  
   Here we aim to place strong lensing on a firmly objective basis
  with the development of a practical non-parametric method for
  inverting the strong lensing image information to extract reliable
  projected 2D surface mass distributions. With the dramatic
  improvement in strong lensing data we can now focus on extracting
  the important physical information with minimal assumptions, in the
  most model independent way, in particular to relax the conventional
  assumption that mass traces light, enabling us to derive the general
  matter distribution and its realistic uncertainties.  These new
  images from Hubble, particularly from the dedicated CLASH program
  \citep{Postman2011}, provide typically over several
  tens of multiply lensed images per cluster and many long arcs, that
  should make this a manageable task. A
  non-parametric approach will provide an important consistency check
  of the findings of the parametric methods since concurring results
  would strengthen the validity of the parametric approach, whereas
  any significant differences would need to be addressed.

    To date, non-parametric methods have been applied to only three well
  studied clusters, using a modification of the strong lensing package
  developed originally by \citep{Diego2005b}, providing low resolution
  representations of the mass distributions and the very different
  non-linear approach of Liesenbourgs et al. \citep{Liesenborgs2006}, applied to the Hubble data
  of Cl0024 \citep{Zitrin2009}. These methods are able to provide the rough
  shape of the mass distributions, showing substructure that roughly
  coincides with clumps in the galaxy distribution, as well as
  reasonably accurate radial mass profiles that are consistent with
  our standard parametric modelling \citep{Diego2005a, Zitrin2009}. It is also clear that this approach cannot help find new
  multiple images, because of its limited resolution, and relies on
  the input multiple images defined by parametric model of \citep{Broadhurst2005,  Zitrin2009}
  and needs reliable
  redshift information for these systems for a meaningful constraint
  on the gradient of the mass profile. Some degeneracies are also
  present, including possible spurious ring features, probably caused by over fitting the
  data \citep{Ponente2011}, and a tendency to asymptote to a flat
  outer profile beyond the boundary of the data, from the mass-sheet
  degeneracy \citep{Jee2007}.

  The results obtained with parametric and non-parametric methods are
 not expected to agree in detail, as the premise on which the
 parametric models are built use optical based information rather than
 the invisible dark matter. Typically parametric models place halos of
 matter coincident with the location of a brightest cluster galaxy,
 with other sub halos added to help deal with any obvious substructure
 seen near the cluster center. For every halo added at least 6
 parameters are required to describe the halo position, ellipticity
 position angle, scale length and profile slope. With additional
 parameters describing cluster member galaxies resulting typically
 in many parameters of uncertain validity, requiring many
 multiply lensed images to constrain. This is particularly the case 
 for ongoing merging clusters. 

% COMMENTED BY CHEMA. THIS PARAGRAPH IS REDUNDANT. T IS SAID (BETTER) ABOVE.
% In cases where different solutions are obtained from the same data
% set, it is important to identify and understand where these
% differences come from.  In earlier works we identified some of the
% limitations of non-parametric methods. One of the most relevant ones
% was the relatively poor resolution of the solution when compared with
% the parametric methods. The non-parametric solutions often render
% solutions that appear as smoother versions of the parametric methods
% (in those cases where the different methods produce similar
% solutions). When applied over simulated data, the non-parametric
% solution is again a smoother representation of the true mass. Also,
% connected with this fact, non-parametric solutions are renormalized
% in order for the solution not to over-fit the data (see the discussion
% about the point source solution in \citep{Diego2005a}).

 In this paper we augment the earlier non-parametric code, WSLAP, by
 incorporating the lens deflection generated by observed member
 galaxy properties, which it transpires helps solve some of the issues
 of non-parametric methods and greatly improves the quality and robustness of the mass
 reconstruction. We do this by including a physical prior in the
 method that is well motivated by the observations. Our prior consist
 in the simple assumption that the galaxies that are in the cluster
 must have some mass themselves and that they are surrounded by their
 own halo of dark matter. In the Sec. \ref{method}   below we discuss the basis
 of the original code, WSLAP, and show how to include the above
 physical prior in the improved version of the code, WSLAP+. In
Sec. \ref{simulation} we give details of the simulations being used to demonstrate the
 capability of the new version of the code, WSLAP+, and finally we describe the results we obtain 
 and our conclusions in Sec. \ref{results} and \ref{conclusions}.
  
%%%%%%%%%%%%%%%%%%%%%%%%%%%%%%%%%%%%%%%%%%%%%%%%%%%%%%%%%%%%%%%%  
\section{The original code: WSLAP}  
\label{method}  
%%%%%%%%%%%%%%%%%%%%%%%%%%%%%%%%%%%%%%%%%%%%%%%%%%%%%%%%%%%%%%%%  
We refer the reader to the original papers \citep{Diego2005a, Diego2005b, Diego2007} for
a detailed description of the original code and its performance. Here we will
summarize the main ideas and those that are relevant to understand the
new improvement to the original method (i.e the addition of a new physical prior).

Gravitational lensing is formally described by the lens equation:
\begin{equation}  
\vec{\theta} = \vec{\beta} + \vec{\alpha}(\vec{\theta},M(\vec{\theta})).  
\label{eq_lens}  
\end{equation}  
In the context of the thin lens approximation, the above equation relates  
the observed lensed images, $\vec{\theta}$, in the image plane (and represented by 
$N_{\theta}$ pixels in the image data) with the corresponding original positions of the 
background galaxies, $\vec{\beta}$, in the source plane  
and the deflection due to the mass distribution, $\vec{\alpha}(\vec{\theta},M)$, in the lens plane.  
For a given mass distribution, $M(\vec{\theta}$, the net deflection angle due to this mass 
is the integral of the deflection field from the infinitesimal mass elements, 
\begin{equation}  
\alpha(\theta) = \frac{4G}{c^2}\frac{D_{ls}}{D_s D_l} \int M(\theta')  
                 \frac{(\theta - \theta')}{|\theta - \theta'|^2} d\theta' ,
\label{eq_alpha}  
\end{equation}  
where $D_{ls}$, $D_l$, and $D_s$ are the angular distances from the
lens to the source, the observer to the lens and from the
observer to the source respectively.

If the lens plane is discretized into a 2-dimensional grid with $N_c$ grid points, 
the above equation can be {\it approximated} as,
\begin{equation}  
\alpha(\theta) = \frac{4G}{c^2}\frac{D_{ls}}{D_s D_l} \sum_i^{N_c} m_i  
                 \frac{\left(\theta - \theta_i\right)}{|\theta - \theta_i|^2},  
\label{eq_alpha2}  
\end{equation}  
where $m_i$ are the masses from each grid point. 
As detailed in previous papers \citep{Diego2005a, Diego2005b, Diego2007} the  
masses at the grid points are modelled as Gaussian with a full-with-half-maximum proportional 
to the mesh size of the grid. 

It is important to emphasize that Eq. \ref{eq_alpha2} represents an approximation of 
Eq. \ref{eq_alpha} and that as such we are introducing an error in the reconstruction. This 
intrinsic error is not always acknowledged in lensing reconstruction and can lead to erroneous 
conclusions as discussed in \citep{Ponente2011}. 

 A second approximation allows us to re-write the lens equation in a simpler algebraic form. 
 Assuming that our data set consists of $N_{\theta}$ lensed pixels of $N_s$ background sources and that 
 each of the $N_s$ is well approximated by a point source (with parameters $\beta_o^x$ and  $\beta_o^y$), 
 we can construct a system of $2N_{\theta}$ (x and y) linear equations with $2N_s+N_c$ variables. 
\begin{equation}  
\left( \begin{array}{c} \vec{\theta}_x \\  
                        \vec{\theta}_y \end{array} \right) =   
\left( \begin{array}{ccc} \hat{\Upsilon}_x \ \ \hat{1} \ \ \hat{0} \\  
                          \hat{\Upsilon}_y \ \ \hat{0} \ \ \hat{1} \end{array} \right)  
\left( \begin{array}{c} \vec{M} \\  
                        \vec{\beta}_o^x\\  
                        \vec{\beta}_o^y\\ \end{array} \right).
\label{eq_lens3}  
\end{equation}  
Here $\hat{\Upsilon}_x$ and $\hat{\Upsilon}_y$ are two $N_{\theta}
\times N_c$ matrices containing the $x$ and $y$ lensing effect of the
cell $j$ (which has been assigned a fixed mass) on the $\theta$ pixel $i$, 
while $\hat{1}$ and $\hat{0}$ are $N_{\theta} \times N_s$ dimensional matrices 
filled with 1's and 0's respectively.

 The variables are the $N_c$ lens masses and 
 the $2N_s$ central galaxy positions (x and y). All these variables can be combined into a single vector, 
 $\vec{X}=(\vec{M},\vec{\beta}_o^x,\vec{\beta}_o^y)$. In its compact form the above equation then reads;
\begin{equation}  
\vec{\Theta} = \Gamma \vec{X},
\label{eq_lens2}  
\end{equation}  
where $\Gamma$ is a known $2N_{\theta}\times (N_c+2N_s)$ dimensional matrix and $\vec{\Theta}$ is also known and given by the 
observed x and y positions of all the pixels in the lensed galaxies.
 
A solution of the system Eq. \ref{eq_lens2} can be found easily by different 
methods (bi-conjugate gradient, singular value decomposition, and quadratic programming (QADP) 
that where already studied by \citep{Diego2005a} but many others can be applied to the same system).

%%%%%%%%%%%%%%%%%%%%%%%%%%%%%%%%%%%%%%%%%%%%%%%%%%%%%%%%%%%%%%%%  
 \section{New implementation: WSLAP+} \label{MethodImproved}
%%%%%%%%%%%%%%%%%%%%%%%%%%%%%%%%%%%%%%%%%%%%%%%%%%%%%%%%%%%%%%%%  

\begin{figure}
\centering
	\includegraphics[scale =0.45] {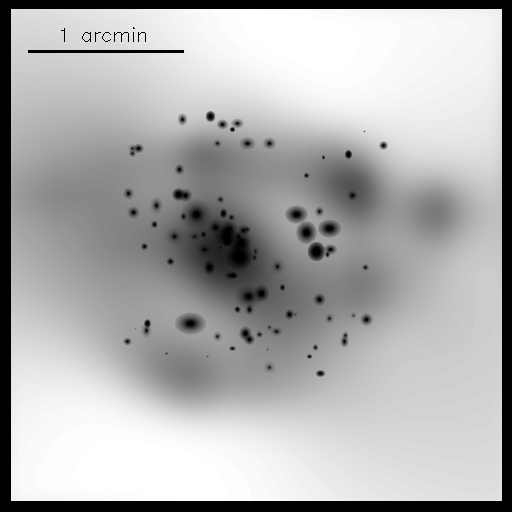}
	  %  \rule{4cm}{3cm}
	\caption{Simulated cluster at $z=0.185$. 
                 The total mass is $2.58\times10^{14} M_\odot/h$ and the field of view is $3.3$ arcminutes across. 
                 In order to better show the matter in the galaxies and in the soft dark matter halo, 
                the galaxies have been saturated and the color scale has been adjusted to increase contrast.}
	\label{fig:gal2dmapTrue}
\end{figure}

\begin{figure}
\centering
\includegraphics[scale =0.45]{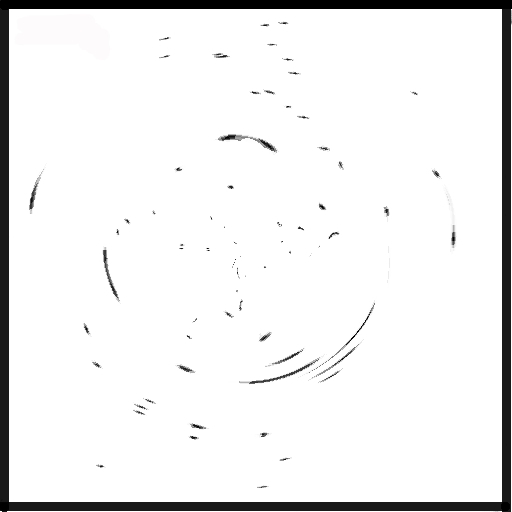}
\caption{Simulated arcs from the mass distribution in Fig. \ref{fig:gal2dmapTrue} and a distribution of simulated sources behind the cluster and at different redshifts.\label{fig:thetaM1} }

\end{figure}

As mentioned earlier, non-parametric methods trade spatial resolution by robustness in the lensing 
reconstruction. On the other hand, parametric methods force matter to concentrate around the observed 
galaxies and usually complement this with a cluster halo described by several parameters. In our new 
implementation, we extend WSLAP by adding a very simple but robust constrain that combines the benefits of 
the robustness from non-parametric methods with the higher resolution of the parametric methods. 
The galaxies in the cluster {\it must} contain some matter and hence they  {\it must} contribute to the deflection 
field. 
%COMMENTED BY CHEMA: REDUNDANT?
%In some cases, the deflection field from individual galaxies competes with the deflection field from 
%the entire cluster (for instance, the surface mass density around prominent cluster members can easily exceed 
%the critical density when the cluster mass is added to the galaxy mass and hence 
%produce lensed images that could bias the lens reconstruction if the effect of the galaxy is not accounted for). 
Due to the intrinsic non-linear nature of the lensing problem, the intrinsically smaller (compared with the cluster) 
deflection field from an individual galaxy in the cluster can make a big difference (sometimes drastic) 
in terms of lensing distortion when the angular distance to this galaxy is small enough. 
Hence it is important to take into account this small deflection angles into the lens reconstruction. 
We can take advantage of the well known correlation between the observed luminosity of a galaxy and its 
total mass and assign a mass to each galaxy in the cluster according to its luminosity.  
As an initial guess we consider a ratio between the luminosity and the mass $\sim 20$. 
Given the mass of a galaxy we assign a NFW mass profile \citep{Navarro1997} to each galaxy . 
We produce a mass map for the haloes around 
the galaxies in the cluster and from this construct a fiducial deflection field for the different 
redshifts of the background sources. 
The deflection field from these galaxies can be easily incorporated into the 
$\Gamma$ matrix, Eq. \ref{eq_lens2}, by adding a column containing the fiducial deflection field at the 
positions of the arcs (lensed galaxies), $\vec{\alpha}_{\rm{gal},x},\vec{\alpha}_{\rm{gal},y}$. 
The new system of equations has the following form:

% \is{($M_{\rm{fid},i}10^{-\rm{mag}_I/2.5+5}$)} \Ch{add reference and/or needs better explanation/motivation. 
% A mag(i)=25 predicts 1.0E5 ? What are the units ?}. 
% Given the mass of a galaxy we can assign a mass profile to that galaxy (for instance a Navarro, Frenk and White 
% profile  \citep{Navarro1997} with a scale radius \is{ $R_{\rm{fid},i}\simeq (10/3.5) M_{\rm{fid}}^{1/3}$)} \Ch{same as above, 
% reference and units? For instance, self-similar models predict $R_v = 1.3*M_{15}^{1/3}$ Mpc/h and $R_{\rm{fid},i}$ should be several 
% times smaller (5-10?) than the virial radius, $R_v$}. 
% Repeating this procedure with each galaxy we can produce a mass map for the haloes around 
% the galaxies in the cluster and from this construct a fiducial deflection field for the different redshifts of the background sources. 
% The deflection field from these galaxies can be easily incorporated into the 
% $\Gamma$ matrix, Eq. \ref{eq_lens2}, by adding a column containing the fiducial deflection field at the 
% positions of the arcs (lensed galaxies), $\vec{\alpha}_{\rm{gal},x},\vec{\alpha}_{\rm{gal},y}$. 
% The new system of equations has the following form:
\begin{equation}  
\left( \begin{array}{c} \vec{\theta}_x \\  
                        \vec{\theta}_y \end{array} \right) =   
\left( \begin{array}{cccc} \hat{\Upsilon}_x \ \ \vec{\alpha}_{\rm{gal},x} \ \ \hat{1} \ \ \hat{0} \\  
                          \hat{\Upsilon}_y \ \ \vec{\alpha}_{\rm{gal},y} \ \ \hat{0} \ \ \hat{1} \end{array} \right)  
\left( \begin{array}{c} \vec{M} \\  
                         C_{\rm{gal}} \\
                        \vec{\beta}_o^x\\  
                        \vec{\beta}_o^y\\ \end{array} \right) . 
\label{eq_lens4}  
\end{equation}  
where  $C_{\rm gal}$ is a new variable (scalar) in the solution vector that accounts for the re-scaling of the fiducial 
deflection field of the galaxies. This system can be also represented in the compact form given by Eq. \ref{eq_lens2}
where now the solution vector $\vec{X}$ is given by $X=(\vec{M},C_{\rm{gal}},\vec{\beta_o^x},\vec{\beta_o^y})$ and it has 
dimension $N_c + 1 + 2N_s$.

As mentioned earlier, Eq. \ref{eq_lens2} can be solved by different methods 
(see \cite{Diego2005a, Diego2005b, Diego2007} for a description of several of them). In our particular case, and in order to 
avoid solutions with negative values in $\vec{M}$ and $C_{\rm{gal}}$ we use the quadratic programming algorithm (or QADP) described 
in \citep{Diego2005a} which imposes the physical constraint that the solution, $\vec{X}$, must be positive. 

Although not discussed in detail in this work, the original code combines also weak lensing (when available) into a 
system of linear equations similar to Eq. \ref{eq_lens2}. The new implementation discussed in the following section can be easily 
extended to weak lensing  by inserting additional column(s) into the corresponding weak lensing matrix $\Gamma$ 
(see \cite{Diego2007} for details of this matrix).

%%%%%%%%%%%%%%%%%%%%%%%%%%%%%%%%%%%%%%%%%%%%%%  
\section{Simulated data}  \label{simulation}
%%%%%%%%%%%%%%%%%%%%%%%%%%%%%%%%%%%%%%%%%%%%%%  
We test the performance of our new code with a set of simulated strong lensing data. 
Our simulated data set resembles the case of A1689, where tens of background sources are being 
lensed by the cluster. We adopt the redshift of the (30) background sources from the real data set of A1689 \citep{Broadhurst2005}. 
For the cluster, we place individual triaxial NFW halos at z=0.2 
with a pattern similar to the distribution of the main halos in A1689 (94 NFW halos in total, \cite{Coe2010}). 
From now on, we refer to the mass distribution from these galaxies as {\it galaxy-true}.  
In addition to the masses from the galaxies, we add a cluster halo (also at z=0.2) with a mass distribution that 
resembles the galaxy distribution but with some significant deviations in order to test how well the method can 
reconstruct the dark matter that is not being traced by the galaxies. 
The mass ratio of the cluster halo to the combined mass of the galaxies is roughly 3 to 1.
The resulting mass distribution of the simulated cluster is shown in Fig. \ref{fig:gal2dmapTrue}. 
In the source plane, the background sources are placed in positions such that we reproduce both tangential and radial 
arcs like in A1689. The background sources are extracted from the Hubble Ultra Deep Field \citep{Bouwens2003}
and later re-scaled to match a specific angular scale at the corresponding redshift. The background simulated sources 
are lensed through the simulated cluster and we produce a set of strongly lensed galaxies that constitutes our simulated data set together with 
the redshifts of the corresponding sources (see Fig. \ref{fig:thetaM1}). The field of view of this (and all other images unless mentioned otherwise) 
is $3.3$ arcminutes.

We also simulate a second mass distribution for the galaxies, that we refer from now on to as {\it galaxy-model}, 
where we use the same locations as above but we change the individual mass and scale radius of each galaxy. 
We take random values for both the mass and scale radius around the values in the galaxy-true case with typical 
deviations of $20\%$ around these values. 
The galaxy-model is later used to compute the fiducial deflection field in our lens reconstruction. 
By doing this, we adopt the realistic scenario where the positions of the galaxy members are known but the 
mass and profiles of these galaxies are unknown.  

\begin{figure}
\centering
	\includegraphics[scale =0.45] {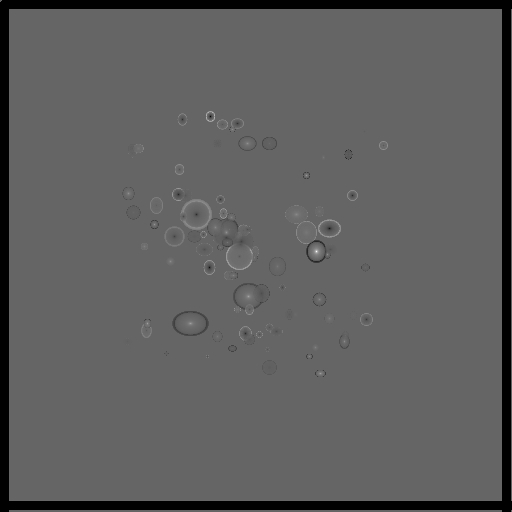}
	  %  \rule{4cm}{3cm}
	\caption{2D map showing the difference in mass between the input model for the galaxies and the model used to reconstruct the cluster mass.}
	\label{fig:gal2dmap}
\end{figure}

\begin{figure}
\centering
	\subfigure[ $\alpha_{gal,x}$ ]{\includegraphics[scale =0.22] {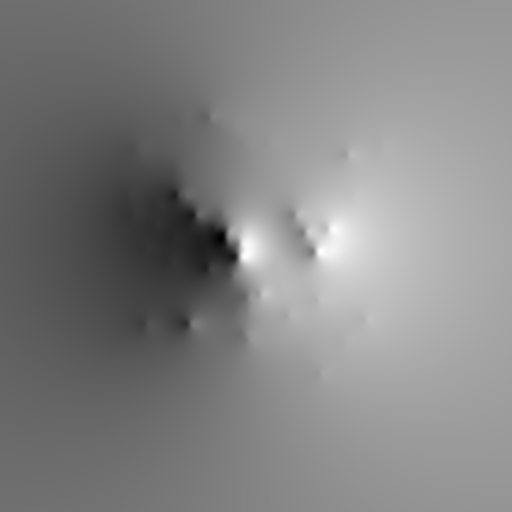}
	    \label{fig:galalphax}
	}
	\subfigure[ $\alpha_{gal,y}$ ]{\includegraphics[scale =0.22] {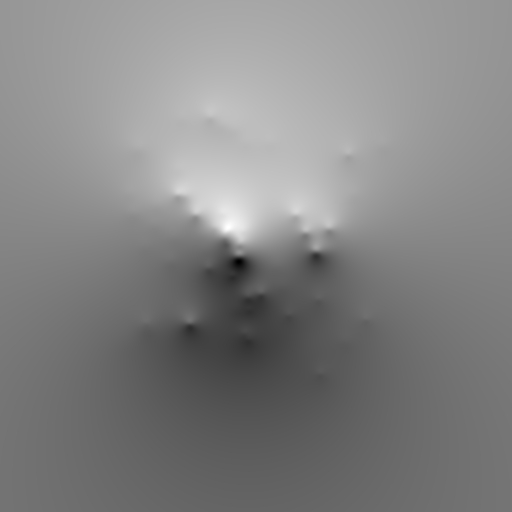} 
	    \label{fig:galalphay}
	}
	\subfigure[ $\alpha_{gal,x}$]{\includegraphics[scale =0.22] {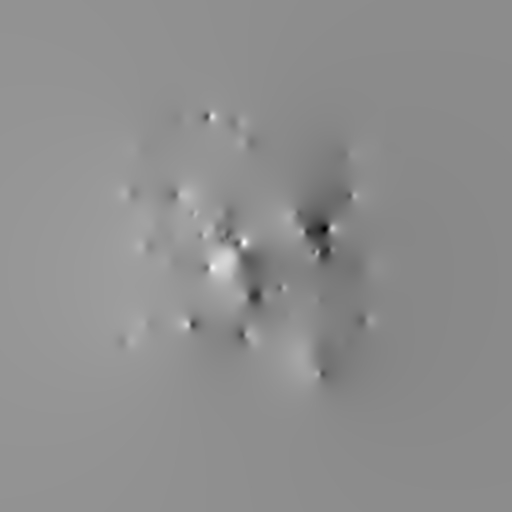}
	    \label{fig:galalphadiffx}
	}
	\subfigure[ $\alpha_{gal,y}$ ]{\includegraphics[scale =0.22] {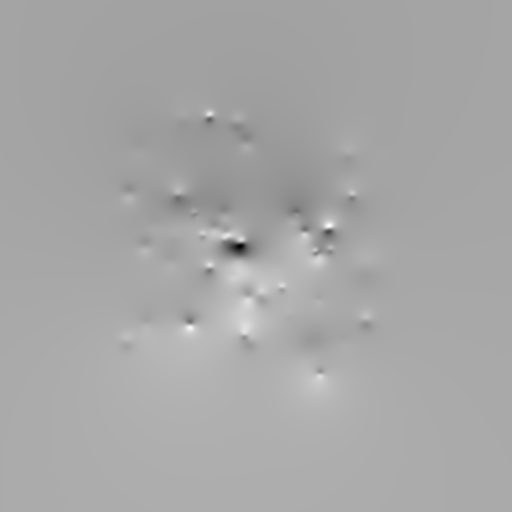}	
	    \label{fig:galalphadiffy}
	}
	\caption{The top panels show the deflection field of the input mass model for the galaxies shown in Fig. \ref{fig:gal2dmap} and the bottom panels show the difference in deflection fields between the input galaxy model and the galaxy-model used for the lensing reconstruction.}
	\label{fig:Galalphamap}
\end{figure}

Fig. \ref{fig:gal2dmap} shows the difference in the projected $2$D surface mass density between the input model distribution of galaxies and the fiducial model used in the mass reconstruction.
The corresponding deflections field are shown in Fig. \ref{fig:Galalphamap}.

Once the fiducial deflection field for the model is computed, we build the $\Gamma$ matrix and reconstruct the solution using 
the QADP algorithm. For the $\Gamma$ matrix we found that using a regular grid (in our case of $32\times32$ grid points) 
works better than a multi-resolution grid. The reason probably being the fact that the multi-resolution grid reduces the 
desired orthogonality of the base (i.e between the grid and  the galaxies) describing the mass distribution. 
Also, the use of a multi-resolution grid can introduce an undesired prior in the reconstruction since the solution tends to artificially 
increase the reconstructed density in the smaller grid cells. 
On the other hand, the use of the regular grid is similar to using a flat prior for the mass distribution since 
it assigns to the different areas in the lens plane the same probability. 
In order to quantify the gain in the reconstruction by the new implementation, we reconstruct the solution in three different scenarios:

$\bullet$   i) Assume that the galaxies in the cluster have zero mass (this would correspond to the result obtained with the 
               original WSLAP code and in general with a standard non-parametric code using a regular grid).\\
$\bullet$  ii) Assume that the mass in the galaxies is given by the galaxy-model and build the fiducial deflection field from 
               that model (Figs.\ref{fig:gal2dmap} and \ref{fig:Galalphamap}). This would be the realistic case where 
               we make an assumption (biased) about the masses in the galaxies. \\
$\bullet$ iii) Assume that the fiducial deflection field is given by the galaxy-true. This case is the best case scenario and 
               corresponds to the best possible reconstruction in the unlikely-lucky case that our assumption about the member 
               galaxies is completely right.

%%%%%%%%%%%%%%%%%%%%%%%%%%%%%%%%%%%%%%%%%%  
\section{Results}  \label{results}
%%%%%%%%%%%%%%%%%%%%%%%%%%%%%%%%%%%%%%%%%%  

\begin{figure*}
\centering
	\subfigure[]{\includegraphics[scale =0.22] {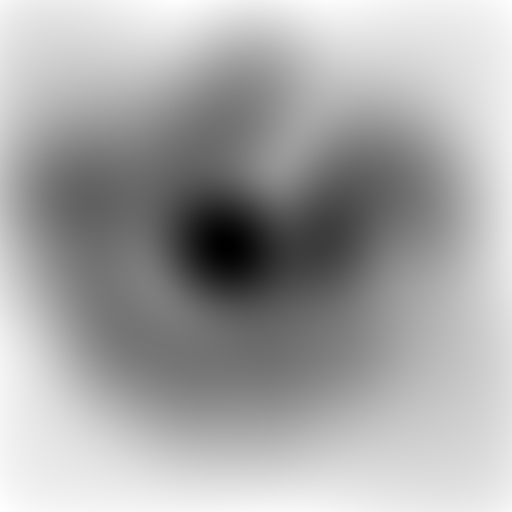}
	  %  \rule{4cm}{3cm}
	    \label{fig:RecompMassNoGal}
	}
	\subfigure[]{\includegraphics[scale =0.22] {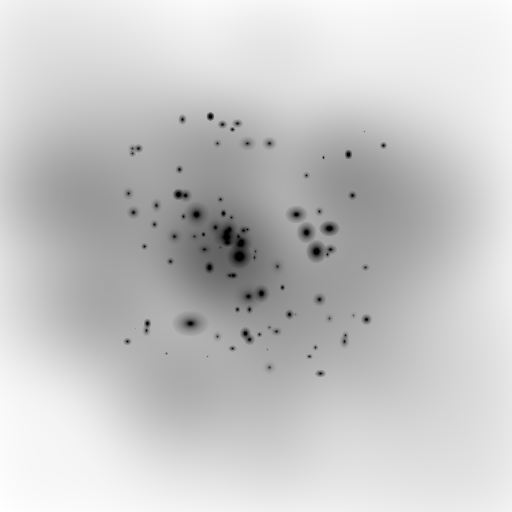}
	  %  \rule{4cm}{3cm}
	    \label{fig:RecompMassModel}
	}
	\subfigure[]{\includegraphics[scale =0.22] {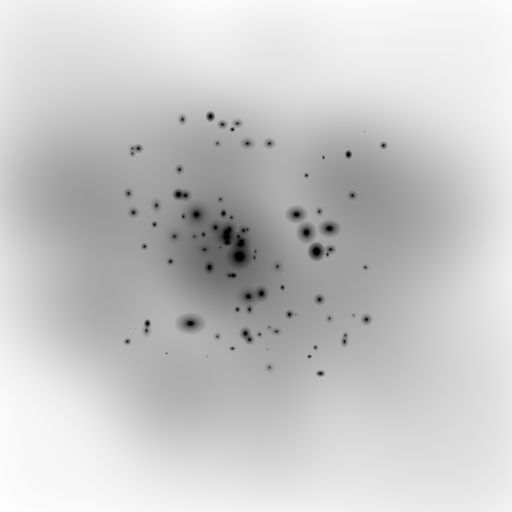}
	  %  \rule{4cm}{3cm}
	    \label{fig:RecompMassCase3}
	}
	\subfigure[]{\includegraphics[scale =0.22] {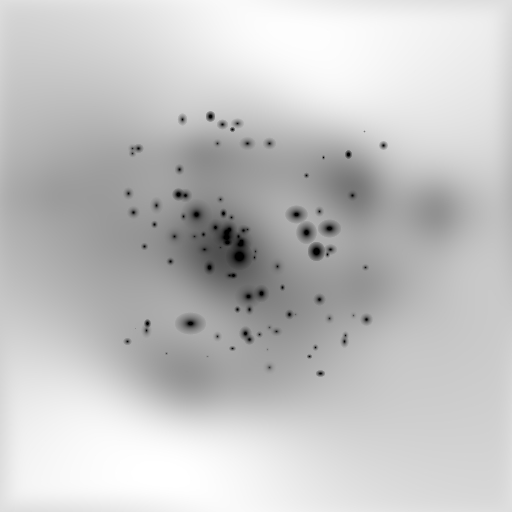}
	  %  \rule{4cm}{3cm}
	    \label{fig:RecompMassTrue}
	}
%	\subfigure[]{\includegraphics[scale =0.22] {Figures/recomposed_critical_curve_Nogalv5fi.jpg}
%	  %  \rule{4cm}{3cm}
%	    \label{fig:CritCurvNoGal}
%	}	
%	\subfigure[]{\includegraphics[scale =0.22] {Figures/recomposed_critical_curve_Modelv5fi.jpg}
%	   % \rule{4cm}{3cm}
%	    \label{fig:CritCurModel}
%	}
%	\subfigure[]{\includegraphics[scale =0.22] {Figures/recomposed_critical_curve_Case3v5fi.jpg}
%	   % \rule{4cm}{3cm}
%	    \label{fig:CritCurCase3}
%	}
%	\subfigure[]{\includegraphics[scale =0.22] {Figures/recomposed_critical_curve_Truev5fi.jpg}
%	    %\rule{4cm}{3cm}
%	    \label{fig:CritCurTrue}
%	}
	\subfigure[]{\includegraphics[scale =0.22] {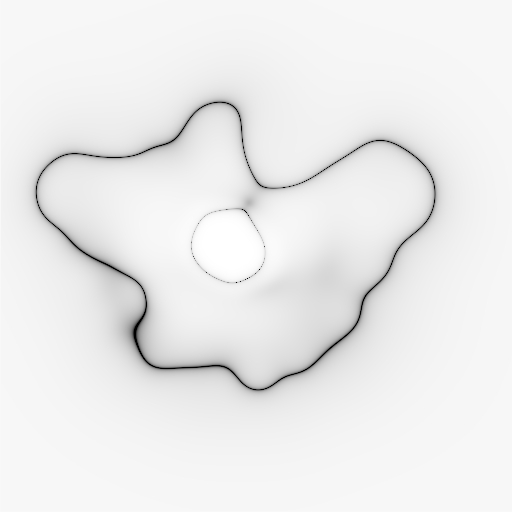}
	  %  \rule{4cm}{3cm}
	    \label{fig:CritCurvNoGal}
	}	
	\subfigure[]{\includegraphics[scale =0.22] {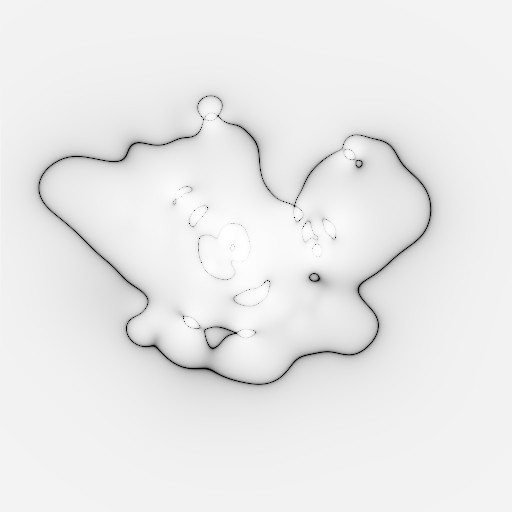}
	   % \rule{4cm}{3cm}
	    \label{fig:CritCurModel}
	}
	\subfigure[]{\includegraphics[scale =0.22] {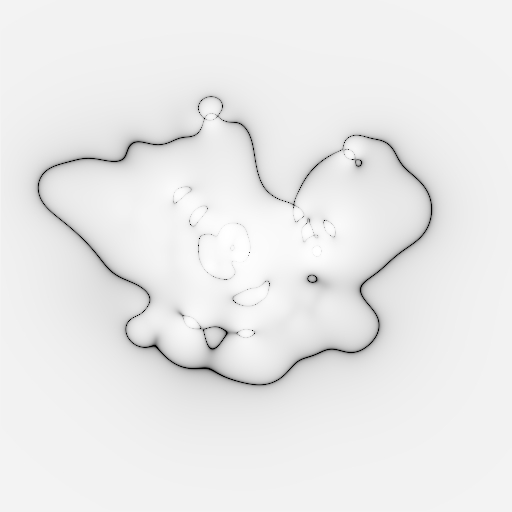}
	   % \rule{4cm}{3cm}
	    \label{fig:CritCurCase3}
	}
	\subfigure[]{\includegraphics[scale =0.22] {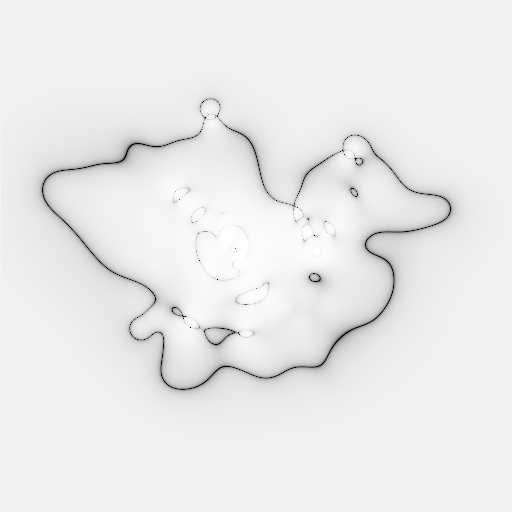}
	    %\rule{4cm}{3cm}
	    \label{fig:CritCurTrue}
	}
	\caption{This figure shows the reconstructed mass map (top row) and critical curve 
    (bottom row) for the different scenarios (case i), ii) and iii)) we have explored and described in Sec. \ref{simulation}   together with the input model  (last column). 
    First column corresponds to case i), while the second and third to
    cases ii) and iii) respectively. 
	\label{fig:Results}}
\end{figure*}

\begin{figure}
\centering
\includegraphics[scale =0.5]{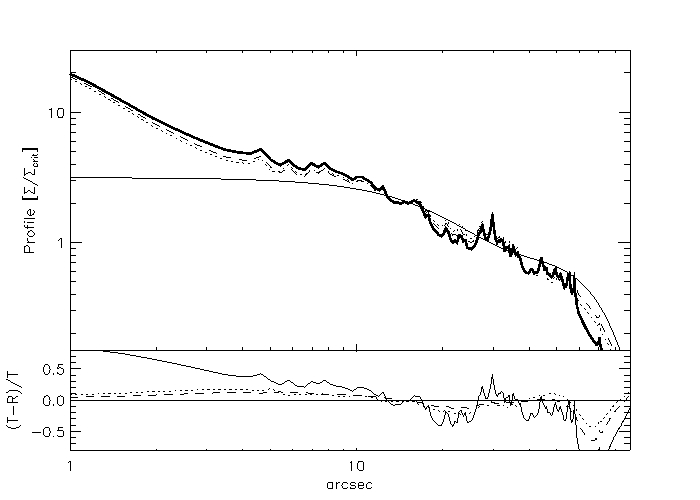}
%\includegraphics[scale =0.5]{Figures/Recomposed1Dcompare_Model_NoGal_True.jpg}
%	\subfigure[Recomposed 1D mass]{\includegraphics[scale =0.33] {Figures/your.jpg}
	    %\rule{4cm}{3cm}
	     \caption{\label{fig:1DMassComp}In the top panel we show the reconstructed profiles. The thick 
solid line corresponds to the input model profile, the thin solid line corresponds to case i) 
(old WSLAP solution), dotted line corresponds to the realistic case ii) and the dashed line 
corresponds to the best case scenario of case iii). The bottom panel shows the the relative differences 
(input model-reconstruction)/input model) between the profiles of cases i), ii) and iii) and the input model profile. The 
line-styles are the same as above.}
\end{figure}

As discussed later, we find that the best solutions are obtained after iterating the QADP for several thousand iterations. 
We find the solution in the 3 cases discussed at the end of the previous section after iterating the QADP algorithm 
for 8000 steps. 
Fig.  \ref{fig:Results} summarizes our main results. Each column corresponds to one of the cases described in the previous section. 
The top row shows the reconstructed mass distribution while the bottom row 
shows the critical curves overlaid the galaxies
Case i) (left column) shows a decent reconstruction of the dark matter halo but as expected misses the details of the individual 
galaxies. This is made more evident when we compare the critical curves with the input model critical curves in the right column. The 
reconstructed critical curves have softer rounds, as a consequence of the poorer resolution of the reconstruction. 
In this case there is only one radial curve. In contrast with the other cases, where the solution is able to reconstruct 
better the critical curves (both radial and tangential). 
%Finally, the de-lensed arcs still show some dispersion in the source plane. 
%Those small deflection angles are the ones that the galaxies in our new 
%implementation can correct better as it is shown in the second and third rows. 
The cases ii) and iii) shown in the second and third columns show a significant improvement in the 
reconstruction of the mass and critical curves. Regarding the mass, it is interesting to see how the grid part of the 
solution is capable of reconstructing the cluster mass structures that where not correlated with the galaxies demonstrating 
the robustness of our new implementation.
On the other hand, the addition of the fiducial deflection field from the galaxies helps improve significantly the recovery of the 
critical curves, in particular the radial critical curve where the effect of the individual galaxies is larger. 
Even the radial curves around the smaller sub-cluster seem to be reconstructed reasonably well.

A more quantitative comparison of the quality of the reconstruction is shown in Fig. \ref{fig:1DMassComp} 
where we compare the one-dimensional profiles (in units of the critical surface density, 
$\Sigma_{crit}=\frac{c^2 D_s}{4\pi G D_l D_{ls}}$) for the three cases and the input model profile. 
Again the new implementation 
is able to reconstruct significantly better the smaller details of the mass distribution. 

\begin{figure}
\centering
\includegraphics[scale =0.23]{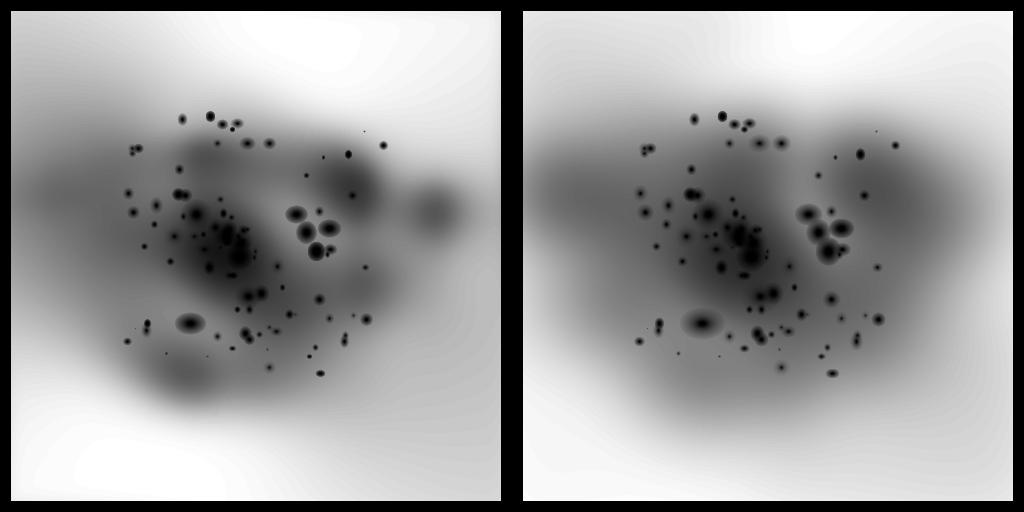}
	   \caption{\label{fig:SoftCompOld}Input model (left) versus reconstructed mass (right) using a color scale that shows better the 
                                               diffuse dark matter component. For comparison purposes, both images are presented in the same 
                                               scale and the galaxies have been saturated to the same value. The recovered distribution follows well the input distribution including the relatively dark substructures that do not trace the input galaxy distribution, and with limiting resolution given by the surface density of the lensed images.}
\end{figure}

%\begin{figure}
%\centering
%\includegraphics[scale =0.2]{Figures/recomposedmassSoftTruevsCase2.jpg}
%%	\subfigure[Recomposed z mass]{\includegraphics[scale =0.33] {Figures/your.jpg}
%	    %\rule{4cm}{3cm}
%	    
%            \caption{\tom{pls elaborate on what this is and add a conclusion type remark to guide the reader..} \is{BETTER NOW? (Left) True versus (Right) reconstruced soft mass component of the cluster. The True one corresponds simulated cluster and it has been used to  obtain
%the arcs, Fig. \ref{fig:thetaM1}. The reconstructed one belongs to case ii) when we use a null galaxy deflection map.}\label{fig:SoftComp} }
%\end{figure}

In order to show the capability of the method to reconstruct dark matter substructure not correlated 
with the galaxies, Fig. \ref{fig:SoftCompOld} compares the reconstructed solution with the input model but using a different color scale that enhances the details  of the soft dark matter halo. In these figures it can be appreciated how the solution retains the main features of this halo although it misses some of 
the details specially near the edges where the lensing constraints are weaker.

%%%%%%%%%%%%%%%%%%%%%%%%%%%%%%%%%%%%%%%  
\section{Discussion}  \label{discusion}
%%%%%%%%%%%%%%%%%%%%%%%%%%%%%%%%%%%%%%%  
Aside from the improvement in the reconstruction of the solution (masses and source positions) shown by the 
reconstructed profiles and critical curves, 
the addition of the new parameter $C_{\rm{gal}}$ results in two major advantages for the new method. One of 
the pathological behaviours of the original code was that the algorithm can not be left 
converging indefinitely. After several thousand iterations, the lack of resolution of the gridded mass 
distribution is generally compensated by a extremely irregular mass distribution that manages to 
{\it focus} the observed arcs into very small compact regions in the source plane. In previous works 
\citep{Diego2005a, Diego2005b, Diego2007, Ponente2011} this pathological solution is referred as to the 
{\it point source solution}. Adding the deflection field from the galaxies 
naturally incorporates the resolution that the grid is lacking so we should expect some improvement on the 
pathological behaviour of the solution when the number of iterations is too large. In order to check the convergence we 
iterate the QADP algorithm a sufficiently large number of iterations. Also, we explore the dependence of the 
solution on the initial guess, $X_o$, for the minimization process.

\begin{figure}
\centering
\includegraphics[scale =0.5]{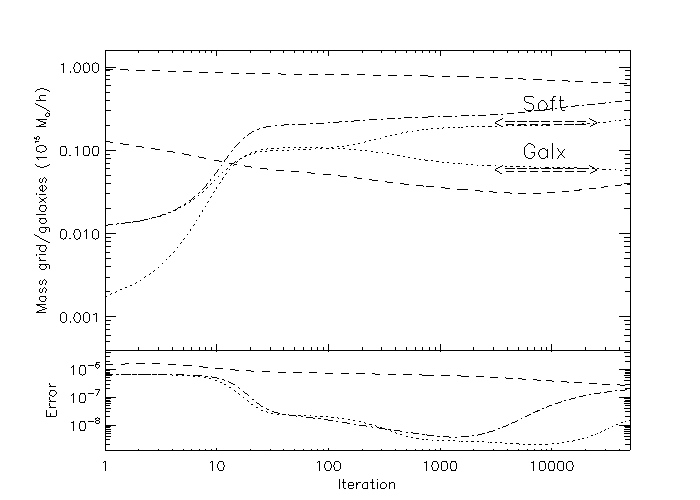}
%	\subfigure[Recomposed 1D mass]{\includegraphics[scale =0.33] {Figures/your.jpg}
	    %\rule{4cm}{3cm}
	    
            \caption{\label{fig:MiterC0C1C2} \textit{Top panel}. Total mass for the grid component and the galaxies component as a 
function of iteration step. The different line styles correspond to the different choices for the initial condition $X_o$ (see text) 
The arrows marked with labels {\it Soft} and 
\textit{Galx} show the  input model total mass of the soft component (dark matter halo) and individual galaxies respectively. 
Note how independently on how good or bad the initial condition is, the solution converges after a few thousand iterations around values 
close to the true values. At around 8000 iterations, the dotted line is almost at the end of a long plateau (optimal solutions are attained in this regime). 
The top dashed line takes longer to converge since it is affected by memory problems in the unconstrained borders of the field of view, 
although in the relevant areas the solution converges towards the input model case as shown by the profiles in Fig. \ref{fig:1DprofileC0C1C2} below. 
For comparison, the dot-dashed line shows the solution obtained by the original WSLAP code (note the overlap with the dotted line in the first few iterations).  
\textit{Bottom panel}. Global error as a function of iteration (see Eq. \ref{eq-err} for a definition of the error). The best solutions 
(excluding the ill-defined dashed line case that fails to converge) are typically obtained after several thousand iterations. 
Beyond many thousand iterations, the solution enters in the over-fitting regime although it still converges to physical solutions 
as shown by the profiles. }
\end{figure}

In Fig. \ref{fig:MiterC0C1C2} we show the total recovered mass of the cluster and the new parameter, $C_{\rm{gal}}$,  
as a function of the iteration number for three different choices of the initial condition $X_o$.  
In the first {\it reasonable} case, (dotted line in the figure), the initial condition has very small values both for the grid 
masses and the $C_{\rm{gal}}$ parameter. 
%In the second {\it optimal} case (dotted line), the initial condition has very small values for 
%the grid masses but the parameter $C_{\rm{gal}}$ is set to a value close to (but not the same) the real input value used to 
%compute the simulated data (arcs). 
In the second {\it bad-choice} case (dashed line) the initial condition is poorly chosen and both 
grid masses and $C_{\rm{gal}}$ are set to values that are too high. For comparison purposes, 
we show a third case (dot-dashed line) with 
the solution for the old WSLAP implementation (or equivalently the acse for $C_{\rm{gal}}=0$)

\begin{figure}
\centering
\includegraphics[scale =0.5]{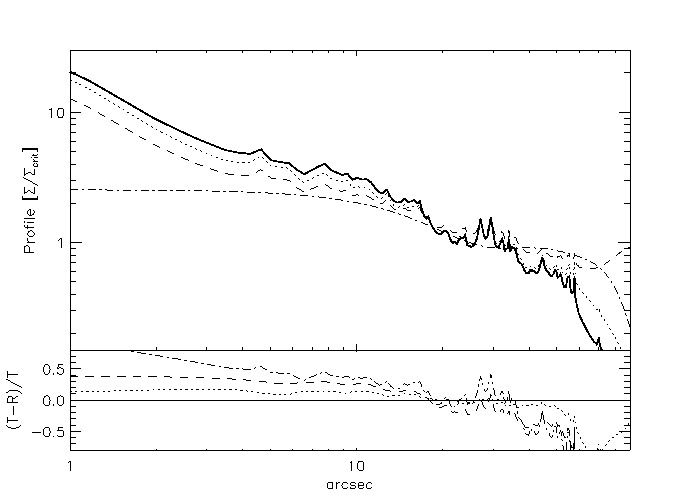}
%	\subfigure[Recomposed 1D mass]{\includegraphics[scale =0.33] {Figures/your.jpg}
	    %\rule{4cm}{3cm}    
\caption{\label{fig:1DprofileC0C1C2}{\it Top panel}. 
Profiles of the input model mass (thick solid line) compared with the profiles of the 
solutions obtained with different initial conditions, $X_o$ (see text) and after 50000 iterations. 
The dotted line shows the case where the initial 
condition has very small values both for the grid masses and the $C_{\rm{gal}}$ parameter,  
%the dotted line shows the case where the initial condition has very small values for 
%the grid masses but the parameter $C_{\rm{gal}}$ is set to a value close to the real mass, 
the dashed line shows the case where the initial condition is poorly chosen and both 
grid masses and $C_{\rm{gal}}$ are set to values that are too high. Note how in this case, the grid suffers 
of memory effects and maintain its initial values at large radii. 
Also shown is the solution obtained by the original WSLAP code (dot-dashed line) after 50000 iterations.
{\it Bottom panel.} 
Relative difference between input model mass (T) and reconstructed masses (R) 
as a function of radius. The different line styles correspond to the same cases 
described above. }
\end{figure}

Despite the choice for $X_o$, after a few thousand iteration steps, the solution ($M$ and $C_{\rm{gal}}$) converges towards constant values. 
Also, these constant values of convergence coincide with the total mass of the diffuse halo of the cluster and the input model mass 
of the galaxies. 
As a difference with the results from the original WSLAP code, this solution is not 
pathological but it is still a good physical solution to the problem. Some degree of over-fitting is still 
appreciated specially in the source plane (where the sources tend to concentrate more towards the center 
of the image) indicating that for this kind of setup (lens, number of arcs, mass distribution) 50000 iterations 
are too many (over-fitting regime) and the optimal range for the number of iterations is around a few thousand (see discussion below). 
The over-fitting regime is better shown in the bottom panel of Fig. \ref{fig:MiterC0C1C2} where we represent the error in the reconstructed 
positions of the sources. 
The error is defined as the absolute difference to the input model solution in terms of separations between 
the input model source positions and the predicted positions 
\begin{equation}
\beta_{err} = \sum(\delta \beta)_x^2 + (\delta \beta)_y^2. 
\label{eq-err}
\end{equation}
Over-fitting usually occurs when the predicted positions of the sources converge towards the 
center of the source plane. This is normally accomplished by non-physical solutions that exhibit large fluctuations in the  
mass distribution. From Fig. \ref{fig:MiterC0C1C2}, we can see that the optimal solutions are obtained in the range of a few thousand iterations. 
The dashed line corresponding to the  {\it bad-choice} described above, fails to converge due to memory effects. 
This memory effect is the reason why the solution does not converge to the {\it Soft} value shown in Fig. \ref{fig:MiterC0C1C2}. 
This is better shown in Fig. \ref{fig:1DprofileC0C1C2} where we compare the profiles of the input model mass with the reconstructed solutions 
in the cases given by the different initial conditions. The solution obtained in the {\it bad-choice} case is still a good one as demonstrated 
by the profile. This figure also shows how the solution maintain the high values (hence memory effect) of the initial condition in the outskirts 
of the image plane, where the lensing data can not constrain the solution. 
The other case seem to render very reasonable solutions even after 50000 iterations (overfitting regime).
%with the {\it optimal} case giving the most satisfactory results. 
For comparison we also show the solution obtained by the original implementation 
of WSLAP (dot-dashed line) and with the same initial guess, $X_o$, as the dotted line ({\it reasonable} case). 
Note how in Fig. \ref{fig:MiterC0C1C2} this case is indistinguishable from the {\it Soft} component of the {\it reasonable} 
choice for the initial guess, $X_o$, for iterations below a thousand but beyond this point it departs from it in a way similar 
to the increase of the {\it Galx} component (bottom dotted line) indicating that the grid is trying to account for the small scale corrections 
due to the member galaxies.
We can then conclude that choosing the optimal number of iterations is not as critical as 
in the original WSLAP code as the over-fitting solutions still are able to reproduce reasonable solutions. However, the best solutions are 
obtained when the number of iterations is in the range of a few thousand. A second lesson is learned about the choice of the initial condition. 
Although the solution is robust and converges towards good-quality solutions independently of the choice for $X_o$, the best solutions are 
obtained when a sensible choice is made for the initial condition, in particular, selecting small values for both, the grid component and the initial strength of the deflection field of the galaxies, produce better final solutions than taking more unreasonable choices.

Figures \ref{fig:MiterC0C1C2} and \ref{fig:1DprofileC0C1C2} summarize some of the main improvements obtained as a result 
of our new implementation. Since the galaxies form on the peaks of the dark matter sub-halos, the galaxy component 
of the solution,  $C_{\rm{gal}}$, will capture the details of the small scale deflection field. The grid 
component, that normally accounts for most of the deflection field, does not need any more to force the mass 
distribution into non-physical solutions (like the dot-dashed line in Fig. \ref{fig:1DprofileC0C1C2} that exhibits a bump or ring of matter at around $1$ arcminute from the cluster centre.) 
to account for the second order corrections to the deflection field 
coming from the smaller halos. This is now naturally accounted for by the galaxy deflection field and hence 
the mass distribution converges to a much more physical (and stable) solution. This pathological behaviour is solved in other  
methods by adding regularization terms. In this sense we can say that our new method produces 
robust self-regularizing solutions where the small scale contributions to the deflection field are described 
by the galaxy component and the irregular (and harder to model) cluster mass distribution is described by the 
grid component. 

\begin{figure}
\centering
\includegraphics[scale =0.45]{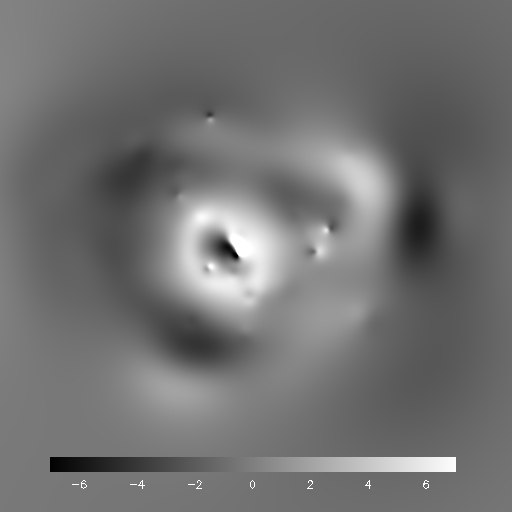}
	    
            \caption{\label{fig:ErrAlpha}Error of the reconstructed deflection field. The units are 
                     arcsecond and they correspond to the difference 
                     $err = \vec{\alpha}_t -\vec{\alpha}_r$ where $\vec{\alpha}_t$ is the 
                     deflection field of the input model and $\vec{\alpha}_r$ is the reconstructed deflection field.}
\end{figure}

A second major bonus is also obtained by incorporating a deflection field for the galaxies with a new free 
parameter. One of the main limitations of the old non-parametric method was the lack of resolution in the 
reconstructed solution. This limitation of the solution made it very difficult to identify new pairs of 
arcs in the images as the error in the deflection field could be large specially around the cluster members. 
This error gets reduced with the new method making the new non-parametric method competitive with the 
parametric methods in terms of finding new arcs in the image. Fig. \ref{fig:ErrAlpha} shows the error in the 
deflection field obtained by comparing the input model deflection field of our simulated data with the deflection 
field of our solution after 8000 iterations. 
%Interestingly, this error is very small near the center of the dominant individual galaxies in the cluster. 
The typical error is about 3 arc-seconds which might be sufficient to identify 
new multiple image-pairs in the data. The largest error is found around the most massive central galaxy, probably 
as a consequence of the wrong assumption made to model the galaxies when computing the fiducial deflection field from 
the galaxies. 

%\begin{figure}
%\centering
%\includegraphics[scale =0.2]{Figures/Critical_Curve_DMHaloMar6v2.jpg}
%%	\subfigure[Recomposed 1D mass]{\includegraphics[scale =0.33] {Figures/your.jpg}
%	    %\rule{4cm}{3cm}    
%\caption{\label{fig:DMhalocheck} {\it Top panel}. 
%True (left) versus reconstructed (right) 2D mass map and critical curves when we add an extra soft dark matter halo (cross)}
%\end{figure}

\subsection{Extension to weak lensing analysis}
%%%%%%%%%%%%%%%%%%%%%%%%%%%%%%%%%%%%%%%%%%%%%%%%%
In the present work we have applied the new improved code, WSLAP+, to simulated strong lensing data. 
The code is however prepared to combine weak and strong lensing as well as detailed in \citep{Diego2005b}. 
The weak and strong lensing data are combined into the same system of linear equations. The same solution (mass 
distribution of the lens) that is able to reproduce the strongly lensed galaxies must predict the right shear 
distortions. The implementation of the weak lensing case in WSLAP+ is the same as the one described in Sec. 
\ref{MethodImproved} for the strong lensing. Now the column containing the deflections from the cluster members 
is extended to include the deflection at the positions where the shear is measured. With our new 
implementation, the small deflection field of a single cluster member (that is, in the outskirts of the 
cluster and can compete in magnitude with the weak lensing shear in the vicinity of that isolated cluster 
member) can be properly accounted for reducing the possible source of systematic error in the weak lensing 
reconstruction. 

\subsection{Adding more than one galaxy deflection field}
%%%%%%%%%%%%%%%%%%%%%%%%%%%%%%%%%%%%%%%%%%%%%%%%%%%%%%%%%%
This paper presents the most simple version of the new implementation where the deflection field from the 
galaxies are described by a model deflection field that is re-scaled by a single parameter, $C_{\rm gal}$. 
It is however trivial to extend this idea to multiple deflection fields. For instance, one might want to 
consider the deflection field from the central galaxy independently. In this case, the $\Gamma$ matrix would 
have two additional columns (with respect to the WSLAP implementation) instead of one and the solution vector, 
$X$, would have two additional free parameters (instead of one), $C_{\rm gal}^1$  $C_{\rm gal}^2$. In a more extreme 
case, the dominant galaxies in the cluster could contribute each with one extra column in the $\Gamma$ matrix 
and their corresponding  $C_{\rm gal}^i$ parameter in the  vector $X$. 
For the case of weak lensing in field areas, this flexibility on the number of parameters might be a necessity 
rather than a convenience since one would normally want to divide the data (lensing galaxies) in redshift bins 
and group the field galaxies into each redshift bin in order to construct a global deflection field for that 
particular redshift bin. This way the number of additional columns in the  $\Gamma$ matrix (and the additional 
number of free parameters in the vector $X$) would be equal to the number of redshift bins that are being 
considered. Incorporating the individual deflection fields from observed galaxies might help improve 
significantly the lensing reconstruction with our new method as the bulk of the dark matter can be well 
described by the grid component but the smaller scale deflection fields around the lensing galaxies 
(that can not be well reconstructed  by the grid) can now be constrained more accurately with the 
individual galaxies deflection field. These and other ideas will be tested in a future paper. 

%%%%%%%%%%%%%%%%%%%%  
\section{Conclusions}  \label{conclusions}
%%%%%%%%%%%%%%%%%%%%  

  We have aimed here to cure the wide degeneracy of lensing solutions
  typical of non-parametric lensing solutions, stressing the
  improvements obtained by treating the cluster member contribution with
  a simple prior.  Cluster members frustrate the process of converging
  to an accurate solution by their small scale perturbations to the
  deflection angle and the additional images they generate. In
  practise it is typically the case that at least one member of a
  multiply lensed source is affected locally in this way by the close
  proximity of a cluster member to the observed image position. We run
  into a limitation here in trying to recover the mass distributions
  of clusters that the effective resolution of the recovered mass maps
  are set by the numbers of lensed images found in the strong lensing
  region, and in practice this is too few to deal with the high
  frequency member galaxy component. In turn this means that
  non-parametric methods is the general inability to predict the
  locations of counter-images with sufficient precision to actually
  find sets of multiple images for adding to the model.
 
 We have found here that this weakness can be largely overcomed by
 incorporating reasonable estimates of the member galaxy deflections
 using the member galaxy positions and luminosity scaled masses, so
 that it then becomes possible to derive the smooth cluster-wide
 component of the mass distribution, for which the variation varies
 only an a relatively large angular scale lying within the effective
 resolution set by the surface density of lensed images. We have
 simply assumed that these galaxies contribute with a mass
 proportional to a fiducial value related to their measured
 luminosities, with the proportionality constant subsequently inferred
 as part of the method. This helps take care of the difficult high
 spatial frequency component, so that the smoother remainder can be
 dealt with by the inherently low resolution non-parametric
 approach. This cluster-wide contribution is modelled with a Gaussian
 pixel grid, providing a compact orthogonal basis. The input data
 includes the multiple images identified by our standard flexible
 parametric model described above, and their redshifts defined from
 our multi-band photometry. By insisting that some mass must exist at
 the position of the observed galaxies we increase the detail of the
 overall reconstruction and also correct possible biases in the
 reconstructed solution as this new assumption can act as an overall
 re-normalization factor.

  Our new method is parameter-free in terms of its description of the
  general cluster mass distribution, so that any interesting anomalous
  density peaks will not go unnoticed in this model-independent
  analysis. Previous work has relied on the assumption that light
  traces mass, and even though we have always tried to relax this
  assumption in our work, it cannot be said that our results in the
  strong lensing regime have much freedom to differ significantly from
  strict equality between mass and light. With our new method we may
  look for deviations between mass and light predicted in the very
  cold Bose-Einstein Condensate (BEC) dark matter, which in contrast to
  standard particle CDM fluctuates in density and may show solitonic
  behaviour and macroscopic interference effects. It is worth noting
  that interesting systematic shifts in position between model images
  and the data of several arc-seconds are quite typical \citep{Broadhurst2005, Halkola2006} which remain
  intriguing. Observationally, we will search for the predicted
  wave-like effects from BEC simulations, and with application to the
  CLASH program.

 We also examine the ability of this method to recover dark
 sub-components which do not follow the galaxy distribution,
 highlighting the potential of this method to uncover such anomalies,
 and for which parametrised models based on the galaxy distribution
 are insensitive.
%, see Fig. \ref{fig:DMhalocheck} from Sec. \ref{discusion}. 
%\is{Check text here. Should we  talk more about the test
%we have done??}\tom{ we could just refer to section X.X}. 

  Finally our new hybrid method has shown that we may be optimistic is
  achieving the precision required to locate multiple images ourselves
  without reliance on other methods to provide the input images. This
  is a major step forward and means that solutions we find by our
  non-parametric technique are self-consistent, in that the multiple
  image we input are derived by our method, and do not need to rely on
  uncertain ``candidates'' which may not be securely identified by
  more model-dependent means. Having derived objective lens models we
  may test the validity of multiply lensed candidates found by others
  and we may also constrain the geometric distances for such
  multiply-lensed sources, and their intrinsic properties, including
  luminosities and source plane reconstructions. This is of particular
  interest in relation to record breaking high-z galaxies routinely
  uncovered in deep cluster imaging, and of potentially great
  importance for the study of structure formation, for which good lens
  models with correspondingly reliable magnification estimates are
  essential. 

  Our first self-consistent application of this technique to the
  iconic cluster A1689 including the new deep IR imaging by Hubble
  will be presented shortly, demonstrating this breakthrough in
  precision by our new non-parametric method allowing new systems to
  be discovered and objective evaluation of the the previously claimed
  multiple images and also a model-independent derivation of lensing
  distances for construction of the distance-redshift relation at high
  redshift. We can anticipate that the most rewarding application
  will be to the newly approved deep ``Frontier fields'' clusters with
  Hubble\footnote{http://www.stsci.edu/hst/campaigns/frontier-fields/} for
  which the high surface density of multiply lensed images strongly
  motivates the objective non-parametric approach to fully explore the
  central surface mass distribution and to reliably estimate the
  magnification of a statistical sample of $z\sim10$ galaxies and beyond.

%%%%%%%%%%%%%%%%%%%%%%%%%%%  
\section{Acknowledgments}  
%%%%%%%%%%%%%%%%%%%%%%%%%%%  
J.M.D., I.S. and R.L. acknowledge support of the Consolider project CSD2010-00064 funded by the  
Ministerio de Economia y Competitividad. I.S. and R.L. are also supported by the Basque Government through the special research action KATEA and
ETORKOSMO, and by the University of the Basque Country UPV/EHU under program UFI 11/55. I.S. also
holded a PhD FPI fellowship contract from the mentioned
ministry at the beginning of this work. T.J.B. thanks
Tzihong Chiueh for interesting
discussions and  J.M.D. and T.J.B. thank
the ASIAA Institute for astronomy in Taipei for their generous
hospitality.

%%%%%%%%%%%%%%%%%%%%%%%%%%%%%%%%%%%%%%%%%%%%%%%%%%%%%%%%%%%%%%%%%%%%%%
\label{lastpage}  
\bibliographystyle{mn2e}
\bibliography{tesis2}

\begin{thebibliography}{}

\bibitem[\protect\citeauthoryear{Bartelmann, Narayan, Seitz \&
  Schneider}{Bartelmann et~al.}{1996}]{Bartelmann1996}
Bartelmann M.,  Narayan R.,  Seitz S.,    Schneider P.,  1996, Astrophys. J.
  Lett., 464, L115

\bibitem[\protect\citeauthoryear{{Bhattacharya}, {Habib}, {Heitmann} \&
  {Vikhlinin}}{{Bhattacharya} et~al.}{2011}]{Bhattacharya2011}
{Bhattacharya} S.,  {Habib} S.,  {Heitmann} K.,    {Vikhlinin} A.,  2011, ArXiv
  e-prints

\bibitem[\protect\citeauthoryear{Boehmer \& Harko}{Boehmer \&
  Harko}{2007}]{Boehmer2007}
Boehmer C.,  Harko T.,  2007, JCAP, 0706, 025

\bibitem[\protect\citeauthoryear{Bouwens, Illingworth, Rosati, Lidman,
  Broadhurst et~al.,}{Bouwens et~al.}{2003}]{Bouwens2003}
Bouwens R.,  Illingworth G.~D.,  Rosati P.,  Lidman C.,  Broadhurst T.~J.,
  et~al., 2003, Astrophys.J., 595, 589

\bibitem[\protect\citeauthoryear{Bridle, Hobson, Lasenby \& Saunders}{Bridle
  et~al.}{1998}]{Bridle1998}
Bridle S.~L.,  Hobson M.~P.,  Lasenby A.~N.,    Saunders R.,  1998, Mon. Not.
  Roy. Astron. Soc., 299, 895

\bibitem[\protect\citeauthoryear{Broadhurst et~al.,}{Broadhurst
  et~al.}{2005}]{Broadhurst2005}
Broadhurst T.~J.,  et~al., 2005, Astrophys.J., 619, L143

\bibitem[\protect\citeauthoryear{Broadhurst, Taylor \& Peacock}{Broadhurst
  et~al.}{1995}]{Broadhurst1995}
Broadhurst T.~J.,  Taylor A.,    Peacock J.,  1995, Astrophys.J., 438, 49

\bibitem[\protect\citeauthoryear{Bullock, Kolatt, Sigad, Somerville, Kravtsov
  et~al.,}{Bullock et~al.}{2001}]{Bullock2001}
Bullock J.~S.,  Kolatt T.~S.,  Sigad Y.,  Somerville R.~S.,  Kravtsov A.~V.,
  et~al., 2001, Mon.Not.Roy.Astron.Soc., 321, 559

\bibitem[\protect\citeauthoryear{Choi}{Choi}{2002}]{Choi2002}
Choi D.-I.,  2002, Phys. Rev. A, 66, 063609

\bibitem[\protect\citeauthoryear{Coe, Benítez, Broadhurst \& Moustakas}{Coe
  et~al.}{2010}]{Coe2010}
Coe D.,  Benítez N.,  Broadhurst T.,    Moustakas L.~A.,  2010, The
  Astrophysical Journal, 723, 1678

\bibitem[\protect\citeauthoryear{{Coe} et~al.,}{{Coe}  et~al.}{2011}]{Coe2011}
{Coe} D.~A.,  et~al., 2011, in A. Ast. Soc. Meeting Abstracts 217 Vol.~43,
  {More Powerful than a Speeding ''Bullet''? New HST Images and Analysis of the
  Galaxy Cluster Merger Abell 2744}

\bibitem[\protect\citeauthoryear{Diego, Protopapas, Sandvik \& Tegmark}{Diego
  et~al.}{005a}]{Diego2005a}
Diego J.~M.,  Protopapas P.,  Sandvik H.,    Tegmark M.,  2005a, Mon. Not. Roy.
  Astron. Soc., 360, 477

\bibitem[\protect\citeauthoryear{Diego, Sandvik, Protopapas, Tegmark, Benitez
  et~al.,}{Diego et~al.}{005b}]{Diego2005b}
Diego J.~M.,  Sandvik H.,  Protopapas P.,  Tegmark M.,  Benitez N.,    et~al.,
  2005b, Mon. Not. Roy. Astron. Soc., 362, 1247

\bibitem[\protect\citeauthoryear{Diego, Tegmark, Protopapas \& Sandvik}{Diego
  et~al.}{2007}]{Diego2007}
Diego J.~M.,  Tegmark M.,  Protopapas P.,    Sandvik H.,  2007, Mon. Not. Roy.
  Astron. Soc., 375, 958

\bibitem[\protect\citeauthoryear{Dolag, Bartelmann, Perrotta, Baccigalupi,
  Moscardini et~al.,}{Dolag et~al.}{2004}]{Dolag2004}
Dolag K.,  Bartelmann M.,  Perrotta F.,  Baccigalupi C.,  Moscardini L.,
  et~al., 2004, Astron.Astrophys., 416, 853

\bibitem[\protect\citeauthoryear{Duffy \& van Bibber}{Duffy \& van
  Bibber}{2009}]{Duffy2009}
Duffy L.~D.,  van Bibber K.,  2009, New Journal of Physics, 11, 105008

\bibitem[\protect\citeauthoryear{Eke, Navarro \& Steinmetz}{Eke
  et~al.}{2001}]{Eke2001}
Eke V.~R.,  Navarro J.,    Steinmetz M.,  2001, Astrophys.J., 554, 114

\bibitem[\protect\citeauthoryear{Gonz\'alez \& Guzm\'an}{Gonz\'alez \&
  Guzm\'an}{2011}]{Gonzalez2011}
Gonz\'alez J.~A.,  Guzm\'an F.~S.,  2011, Phys. Rev. D, 83, 103513

\bibitem[\protect\citeauthoryear{Halkola, Seitz \& Pannella}{Halkola
  et~al.}{2006}]{Halkola2006}
Halkola A.,  Seitz S.,    Pannella M.,  2006, Monthly Notices of the Royal
  Astronomical Society, 372, 1425

\bibitem[\protect\citeauthoryear{Hu, Barkana \& Gruzinov}{Hu
  et~al.}{2000}]{Hu2000}
Hu W.,  Barkana R.,    Gruzinov A.,  2000, Phys.Rev.Lett., 85, 1158

\bibitem[\protect\citeauthoryear{Jee et~al.,}{Jee  et~al.}{2007}]{Jee2007}
Jee M.~J.,  et~al., 2007, The Astrophysical Journal, 661, 728

\bibitem[\protect\citeauthoryear{Kaiser}{Kaiser}{1995}]{Kaiser1995}
Kaiser N.,  1995, Astrophys.J. Lett., 439, L1

\bibitem[\protect\citeauthoryear{Kaiser \& Squires}{Kaiser \&
  Squires}{1993}]{Kaiser1993}
Kaiser N.,  Squires G.,  1993, Astrophys.J., 404, 441

\bibitem[\protect\citeauthoryear{Liesenborgs, De~Rijcke \&
  Dejonghe}{Liesenborgs et~al.}{2006}]{Liesenborgs2006}
Liesenborgs J.,  De~Rijcke S.,    Dejonghe H.,  2006, Monthly Notices of the
  Royal Astronomical Society, 367, 1209

\bibitem[\protect\citeauthoryear{{Limousin, M.}, {Richard, J.}, {Kneib, J.-P.},
  {Brink, H.}, {Pell\'o, R.}, {Jullo, E.}, {Tu, H.}, {Sommer-Larsen, J.},
  {Egami, E.}, {Michalowski, M. J.}, {Cabanac, R.} \& {Stark, D.
  P.}}{{Limousin, M.} et~al.}{2008}]{LimousinM.2008}
{Limousin, M.} {Richard, J.} {Kneib, J.-P.} {Brink, H.} {Pell\'o, R.} {Jullo,
  E.} {Tu, H.} {Sommer-Larsen, J.} {Egami, E.} {Michalowski, M. J.} {Cabanac,
  R.}   {Stark, D. P.} 2008, A\&A, 489, 23

\bibitem[\protect\citeauthoryear{Macciò, Dutton \& Van Den~Bosch}{Macciò
  et~al.}{2008}]{Maccio2008}
Macciò A.~V.,  Dutton A.~A.,    Van Den~Bosch F.~C.,  2008, Mon. Not. Roy.
  Astron. Soc., 391, 1940

\bibitem[\protect\citeauthoryear{Markevitch, Gonzalez, Clowe, Vikhlinin,
  Forman, Jones, Murray \& Tucker}{Markevitch et~al.}{2004}]{Markevitch2004}
Markevitch M.,  Gonzalez A.~H.,  Clowe D.,  Vikhlinin A.,  Forman W.,  Jones
  C.,  Murray S.,    Tucker W.,  2004, The Astrophysical Journal, 606, 819

\bibitem[\protect\citeauthoryear{Marshall, Hobson, Gull \& Bridle}{Marshall
  et~al.}{2002}]{Marshall2002}
Marshall P.~J.,  Hobson M.~P.,  Gull S.~F.,    Bridle S.~L.,  2002, Astrophys.
  J. Supplement Series, 335, 1037

\bibitem[\protect\citeauthoryear{{Mastropietro} \& {Burkert}}{{Mastropietro} \&
  {Burkert}}{2008}]{Mastropietro2008}
{Mastropietro} C.,  {Burkert} A.,  2008, Mon. Not. Roy. Astron. Soc., 389, 967

\bibitem[\protect\citeauthoryear{{Merten} et~al.,}{{Merten}
  et~al.}{2011}]{Merten2011}
{Merten} J.,  et~al., 2011, Mon. Not. Roy. Astron. Soc., 417, 333

\bibitem[\protect\citeauthoryear{Navarro, Frenk \& White}{Navarro
  et~al.}{1997}]{Navarro1997}
Navarro J.~F.,  Frenk C.~S.,    White S.~D.,  1997, Astrophys.J., 490, 493

\bibitem[\protect\citeauthoryear{Neto, Gao, Bett, Cole, Navarro et~al.,}{Neto
  et~al.}{2007}]{Neto2007}
Neto A.~F.,  Gao L.,  Bett P.,  Cole S.,  Navarro J.~F.,    et~al., 2007,
  Mon.Not.Roy.Astron.Soc., 381, 1450

\bibitem[\protect\citeauthoryear{Oguri, Hennawi, Gladders, Dahle, Natarajan
  et~al.,}{Oguri et~al.}{2009}]{Oguri2009}
Oguri M.,  Hennawi J.~F.,  Gladders M.~D.,  Dahle H.,  Natarajan P.,    et~al.,
  2009, Astrophys.J., 699, 1038

\bibitem[\protect\citeauthoryear{{Oguri}, {Takada}, {Umetsu} \&
  {Broadhurst}}{{Oguri} et~al.}{2005}]{Oguri2005}
{Oguri} M.,  {Takada} M.,  {Umetsu} K.,    {Broadhurst} T.,  2005, Astrophys.
  J., 632, 841

\bibitem[\protect\citeauthoryear{{Okabe} \& {Umetsu}}{{Okabe} \&
  {Umetsu}}{2008}]{Okabe2008}
{Okabe} N.,  {Umetsu} K.,  2008, PASJ, 60, 345

\bibitem[\protect\citeauthoryear{Peebles}{Peebles}{1984}]{Peebles1984}
Peebles P.,  1984, Astrophys.J., 277, 470

\bibitem[\protect\citeauthoryear{Ponente \& Diego}{Ponente \&
  Diego}{2011}]{Ponente2011}
Ponente P.,  Diego J.,  2011, A\&A, 535, A119

\bibitem[\protect\citeauthoryear{Postman, Coe, Benitez, Bradley, Broadhurst
  et~al.,}{Postman et~al.}{2011}]{Postman2011}
Postman M.,  Coe D.,  Benitez N.,  Bradley L.,  Broadhurst T.,    et~al., 2011,
  A. Astro. Soc.

\bibitem[\protect\citeauthoryear{Schneider}{Schneider}{1994}]{Schneider1994}
Schneider P.,  1994, Astron. \& Astrophys., 302, 639

\bibitem[\protect\citeauthoryear{Schneider \& Seitz}{Schneider \&
  Seitz}{1995}]{Schneider1995}
Schneider P.,  Seitz C.,  1995, Astron. \& Astrophys., 294, 411

\bibitem[\protect\citeauthoryear{Seitz \& Schneider}{Seitz \&
  Schneider}{1995}]{Seitz1995}
Seitz C.,  Schneider P.,  1995, Astron. \& Astrophys., 297, 287

\bibitem[\protect\citeauthoryear{{Sikivie} \& {Yang}}{{Sikivie} \&
  {Yang}}{2009}]{Sikivie2009}
{Sikivie} P.,  {Yang} Q.,  2009, Physical Review Letters, 103, 111301

\bibitem[\protect\citeauthoryear{{Springel} \& {Farrar}}{{Springel} \&
  {Farrar}}{2007}]{springel2007}
{Springel} V.,  {Farrar} G.~R.,  2007, Mon. Not. Roy. Astron. Soc., 380, 911

\bibitem[\protect\citeauthoryear{Taylor, Dye, Broadhurst, Benítez \& van
  Kampen}{Taylor et~al.}{1998}]{Taylor1998}
Taylor A.~N.,  Dye S.,  Broadhurst T.~J.,  Benítez N.,    van Kampen E.,
  1998, Astrophy. J., 501, 539

\bibitem[\protect\citeauthoryear{{Thompson} \& {Nagamine}}{{Thompson} \&
  {Nagamine}}{2012}]{Thompson2012}
{Thompson} R.,  {Nagamine} K.,  2012, Mon. Not. Roy. Astron. Soc., 419, 3560

\bibitem[\protect\citeauthoryear{Tyson, Kochanski \& Dell'Antonio}{Tyson
  et~al.}{1998}]{Tyson1998}
Tyson J.~A.,  Kochanski G.~P.,    Dell'Antonio I.~P.,  1998, Astrophy. J.
  Lett., 498, L107

\bibitem[\protect\citeauthoryear{Umetsu \& Broadhurst}{Umetsu \&
  Broadhurst}{2008}]{Umetsu2008}
Umetsu K.,  Broadhurst T.,  2008, The Astrophysical Journal, 684, 177

\bibitem[\protect\citeauthoryear{{Umetsu}, {Broadhurst}, {Zitrin},
  {Medezinski}, {Coe} \& {Postman}}{{Umetsu} et~al.}{2011}]{Umetsu2011}
{Umetsu} K.,  {Broadhurst} T.,  {Zitrin} A.,  {Medezinski} E.,  {Coe} D.,
  {Postman} M.,  2011, Astrophys. J., 738, 41

\bibitem[\protect\citeauthoryear{{Umetsu}, {Medezinski}, {Broadhurst},
  {Zitrin}, {Okabe}, {Hsieh} \& {Molnar}}{{Umetsu} et~al.}{2010}]{Umetsu2010}
{Umetsu} K.,  {Medezinski} E.,  {Broadhurst} T.,  {Zitrin} A.,  {Okabe} N.,
  {Hsieh} B.-C.,    {Molnar} S.~M.,  2010, Astrophys. J., 714, 1470

\bibitem[\protect\citeauthoryear{Umetsu, Medezinski, Nonino, Merten, Zitrin
  et~al.,}{Umetsu et~al.}{2012}]{Umetsu2012}
Umetsu K.,  Medezinski E.,  Nonino M.,  Merten J.,  Zitrin A.,    et~al., 2012,
  Astrophys.J., 755, 56

\bibitem[\protect\citeauthoryear{{Velten} \& {Wamba}}{{Velten} \&
  {Wamba}}{2012}]{Velten2012}
{Velten} H.,  {Wamba} E.,  2012, Physics Letters B, 709, 1

\bibitem[\protect\citeauthoryear{{Widrow} \& {Kaiser}}{{Widrow} \&
  {Kaiser}}{1993}]{Widrow1993}
{Widrow} L.~M.,  {Kaiser} N.,  1993, Astrophys. J. Letters, 416, L71

\bibitem[\protect\citeauthoryear{Woo \& Chiueh}{Woo \& Chiueh}{2009}]{Woo2009}
Woo T.-P.,  Chiueh T.,  2009, The Astrophysical Journal, 697, 850

\bibitem[\protect\citeauthoryear{Zhao, Jing, Mo \& Börner}{Zhao
  et~al.}{2009}]{Zhao2009}
Zhao D.~H.,  Jing Y.~P.,  Mo H.~J.,    Börner G.,  2009, The Astrophysical
  Journal, 707, 354

\bibitem[\protect\citeauthoryear{Zitrin et~al.,}{Zitrin
  et~al.}{2009}]{Zitrin2009}
Zitrin A.,  et~al., 2009, Mon. Not. Roy. Astron. Soc., 396, 1985

\bibitem[\protect\citeauthoryear{{Zitrin} et~al.,}{{Zitrin}
  et~al.}{2010}]{Zitrin2010}
{Zitrin} A.,  et~al., 2010, Mon. Not. Roy. Astron. Soc., 408, 1916

\bibitem[\protect\citeauthoryear{Zitrin, Meneghetti, Umetsu, Broadhurst,
  Bartelmann et~al.,}{Zitrin et~al.}{2013}]{Zitrin2013}
Zitrin A.,  Meneghetti M.,  Umetsu K.,  Broadhurst T.,  Bartelmann M.,
  et~al., 2013, Astrophys.J., 762, L30

\end{thebibliography}
%%%%%%%%%%%%%%%%%%%%%%%%%%%%%%%%%%%%%%%%%%%%%%%%%%%%%%%%%%%%%%%%%%%%%%%  

\end{document}